\documentclass[twocolumn,showpacs,preprintnumbers,amsmath,amssymb]{revtex4}

\usepackage{graphicx}
\usepackage{dcolumn}
\usepackage{bm}

\begin{document}


\title{Low energy hadronic contribution to the QED vacuum polarization}

\author{H.~Burkhardt}
\email{Helmut.Burkhardt@cern.ch}
\affiliation{%
CERN, AB Departement, CH-1211 Geneva 23,  Switzerland
}%

\author{B.~Pietrzyk}
\affiliation{ Laboratoire de Physique des Particules LAPP,
IN2P3-CNRS, F-74941 Annecy-le-Vieux Cedex, France
}%
\email{pietrzyk@lapp.in2p3.fr}
             \date{June 30, 2005}

\begin{abstract}
Recent improvements in the low energy e$^+$e$^-$ annihilation data
and their influence on the determination of the hadronic
contribution to the running of the QED fine structure constant at
m$_Z$ are discussed. Using CMD-2 and KLOE measurements in the
$\rho$ region we obtain $\Delta\alpha^{(5)}_{\rm had}(s)$ =
0.02758 $\pm$ 0.00035 at s = $m_Z^2$.
\end{abstract}

\pacs{13.85.Lg, 13.66.Jn, 12.15.Lk}
\maketitle

In the year 2001, we published an updated evaluation of the
hadronic contribution to the running of the QED fine structure
constant\,\cite{BurPie2001}, based on a dispersion integral using
a parametrization of the measured cross section of
e$^+$e$^-\rightarrow$ hadrons. We obtained a hadronic contribution
of $\Delta\alpha^{(5)}_{\rm had}(s)$ = 0.02761 $\pm$ 0.00036 at s
= $m_Z^2$.

Our parametrization in the c.m.s. energy region of the $\rho$, the
contribution of the $\pi^+\pi^-$ final state from threshold to 1.8
GeV, was based on a pion form factor parametrization obtained by
the CMD-2 Collaboration which used results of their measurements
in the c.m.s. energy region between 0.61 and 0.96\,GeV at the
VEPP-2M collider\,\cite{CMD2-1999}. The overall uncertainty of the
$\rho$ region integral, including the statistical uncertainty, was
2.3\% (that of $\Gamma_{ee}$ in \cite{CMD2-1999}) in our analysis.

Since then, the CMD-2 collaboration improved the treatment of
radiative corrections twice. An intermediate improvement has
appeared in the published document\,\cite{CMD2-2001} and an
additional improvement has become available in
2004\,\cite{CMD2-2003}. We have concluded that the most recent
CMD-2 results imply only a small change in the estimate of the
hadronic contribution \cite{BurPie2003}.

Recently, the KLOE collaboration \cite{KLOE2004} has measured the
cross section of e$^+$e$^-\rightarrow\pi^+\pi^-$ with high
statistical accuracy in small energy bins using the ``radiative
return" from the $\phi$ resonance to the $\rho$ in the
$\pi^+\pi^-$ mass range between 0.59 and 0.97\,GeV.

We have been repeatedly asked to update our previous analysis and
to comment on and quantify the influence of recent low energy
measurements by KLOE and CMD-2 on our results. We find that 
the actual change turns out to be very small.
Since the change is very small we have decided to
submit this work as a brief report. To the extent that this report is an
update of a previously published article, the choice is not to
unnecessarily repeat the discussion for energy regions which did
not change.

In our 2001 analysis, we used the parametrization of the pion form
factor obtained by the CMD-2 collaboration. The contribution of
the new results on $\Delta\alpha^{(5)}_{\rm had}(m_Z^2)$ is now
obtained by direct integration between measured KLOE and CMD-2
data points separately. For CMD-2 we use the ``bare'' cross
section and for KLOE the pion form factor data. These are
quantities, in which the vacuum polarization corrections have been
removed. The small $\rho$ contribution from lower and higher
energies, not covered by new data, is evaluated as previously
using the CMD-2 parametrization of the pion form factor. We treat
the systematic uncertainties as fully correlated between different
c.m.s. energies within the CMD-2 experiment. For the integration
of the KLOE data, we constructed a covariance matrix based on the
statistical covariance matrix with the addition of fully
correlated systematic uncertainties as provided by the KLOE
collaboration.

The results obtained from the dispersion integration of the KLOE
and CMD-2 data at $m_Z^2$ are in good agreement with each other.
The systematic uncertainty of the CMD-2 integration (0.6\%) is
smaller than the corresponding uncertainty of the KLOE integration
(1.4\%). On the other hand, the statistical uncertainty of the
CMD-2 integration is slightly larger than the systematic one,
while the statistical uncertainty of the KLOE integration is
negligible. The integration results are combined as independent
measurements in the evaluation of the $\rho$ contribution to
$\Delta\alpha^{(5)}_{\rm had}(m_Z^2)$.

We obtain a value of the hadronic contribution to the running of
the QED fine structure constant of $\Delta\alpha^{(5)}_{\rm
had}(s)$ = 0.02758 $\pm$ 0.00035 at s = $m_Z^2$ corresponding to
$1/\alpha^{(5)}(m_Z^2) = 128.940 \pm 0.048$. The value of the
$\rho$ contribution has changed from 0.00350 in \cite{BurPie2001}
to 0.00347 and the relative uncertainty has decreased from 2.3\%
to 0.9\%. The change of the uncertainty corresponds to the change
of precision from the preliminary CMD-2 \cite{CMD2-1999} data to
the combination of published CMD-2 \cite{CMD2-2003} and KLOE
\cite{KLOE2004} data. The change of the value and the uncertainty
of the hadronic contribution to the running of the QED fine
structure constant at $m_Z^2$ is very small. In fact the $\rho$
region contributes to less than 13\% to the dispersion integral
and is known to much better precision than many of the other
energy domains as can be concluded from the
Table~\ref{tab:contrib}, which is the updated version from the
Ref.~\cite{BurPie2001}.
\begin{figure}
\includegraphics[width=8.5cm]{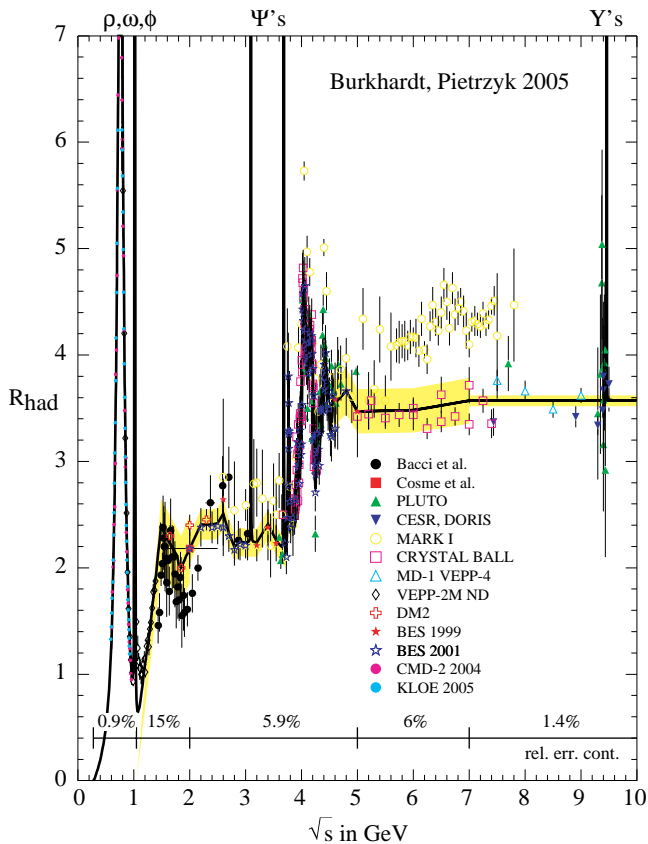}
\caption{\label{plot:RHAD} $R_{\rm had}$ including resonances.
Measurements are shown with statistical errors. The relative
uncertainty assigned to our parametrization is shown as band and
given with numbers at the bottom.}
\end{figure}

We note however, that the shape of the hadronic cross-sections
measured by the KLOE and CMD-2 collaborations differ for some
individual points by more than the systematic uncertainty would
indicate \cite{Beijing}. There also appears to be a small, but
systematic energy shift in the observed cross sections between the
KLOE and the CMD-2  data, which at present is not understood. The
effect on the integrated cross sections which contribute to
$\Delta\alpha^{(5)}_{\rm had}(m_Z^2)$ is however negligible.

The situation is different for the hadronic contribution to the
anomalous  magnetic moment of the muon (g-2)$_\mu$ \cite{DEHZ,
Jegerlehner,Teubner, Davier}. There  the $\rho$ region provides
the dominant contribution. The dispersion  integral involves a
different kernel which gives more weight to lower energies and
larger sensitivity on systematic energy shifts.
\begin{table}[h]
\caption{\label{tab:contrib}Contributions to
$\Delta\alpha^{(5)}_{\rm had}(m^2_Z)$}
\begin{ruledtabular}
\begin{tabular}{ccr}
Range $\sqrt{s}$, GeV& $\Delta\alpha$ & Relative error\\ \hline
$\rho$                 & 0.00347 & 0.9 \% \\
Narrow resonances      & 0.00184 & 3.1 \% \\
1.05 -- 2.0\phantom{0} & 0.00156 & 15 \% \\
2.0 -- 5.0             & 0.00381 & 5.9 \%  \\
5 -- 7                 & 0.00183 & \phantom{0}6 \%  \\
\phantom{0}7 -- 12     & 0.00304 & 1.4 \% \\
\phantom{00}$>$ 12     & 0.01203 & 0.2 \%  \\
\hline
 & 0.02758 & 1.3 \% \\ \hline
\end{tabular}
\end{ruledtabular}
\end{table}

Fig.~\ref{plot:RHAD}, which is the updated version from the
Ref.~\cite{BurPie2001},  gives the summary of $R_{\rm had}$
measurements by different experiments and the current precision in
different e$^+$e$^-$ center-of-mass (cms) energy regions. $R_{\rm
had}$ is the measured QED cross-section of the process
e$^+$e$^-\rightarrow$ hadrons, normalized to the QED cross-section
for lepton-pair production. The uncertainty in the 1-2 GeV energy
region is 15\%. This region contributes to about 40\% to the
uncertainty on dispersion integral at $m_Z^2$, as can be seen from
the Table~\ref{tab:contrib} and Fig.~2 in the
Ref.~\cite{BurPie2001}. We would like to strongly encourage
efforts to measure precisely $R_{\rm had}$ in this cms energy
region.

We would like to thank S.I.~Eidelman from the CMD-2 Collaboration
and A.~Denig and G.~Venanzoni from  the KLOE Collaboration for
useful discussions.

\end{document}